\documentclass[12pt]{article}
\usepackage{amssymb}

\newcommand{\beq}{\begin{eqnarray}}
\newcommand{\eeq}{\end{eqnarray}}
\newcommand{\beqnn}{\begin{eqnarray*}}
\newcommand{\eeqnn}{\end{eqnarray*}}
\newtheorem{theorem}{Theorem}
\newtheorem{lemma}{Lemma}

\newcommand{\proof}{\paragraph*{{\it Proof.\/}}}

\newcommand{\qed}{\fbox{\phantom{-}}\bigskip}
\newcommand{\rd}{\partial}
\renewcommand{\Im}{\mathop{\mathrm{Im}}\nolimits}

\newcommand{\diag}{\mathop{\mathrm{diag}}}
\newcommand{\tp}[1]{\:{}^{\mathrm{t}}#1}
\newcommand{\Ker}{\mathop{\mathrm{Ker}}}
\newcommand{\Coker}{\mathop{\mathrm{Coker}}}

\newcommand{\CC}{\mathbf{C}}

\newcommand{\ZZ}{\mathbf{Z}}
\newcommand{\fkg}{\mathfrak{g}} %%% amssymb
\begin{document}

%%%%%%%%%%%%%%%%
%% title page %% 
%%%%%%%%%%%%%%%%

\title{Landau-Lifshitz hierarchy and \\
infinite dimensional Grassmann variety}
\author{Kanehisa Takasaki\\
{\normalsize Graduate School of Human and Environmental Studies, 
Kyoto University}\\
{\normalsize Yoshida, Sakyo, Kyoto 606-8501, Japan}\\
{\normalsize E-mail: takasaki@math.h.kyoto-u.ac.jp}}
\date{}
\maketitle

\begin{abstract}
The Landau-Lifshitz equation is an example of soliton equations 
with a zero-curvature representation defined on an elliptic curve.  
This equation can be embedded into an integrable hierarchy of evolution 
equations called the Landau-Lifshitz hierarchy.  This paper elucidates 
its status in Sato, Segal and Wilson's universal description of soliton 
equations in the language of an infinite dimensional Grassmann variety.  
To this end, a Grassmann variety is constructed from a vector space of 
$2 \times 2$ matrices of Laurent series of the spectral parameter $z$.   
A special base point $W_0$, called ``vacuum,'' of this Grassmann variety 
is chosen.  This vacuum is ``dressed'' by a Laurent series $\phi(z)$ to 
become a point of the Grassmann variety that corresponds to a general 
solution of the Landau-Lifshitz hierarchy.  The Landau-Lifshitz hierarchy 
is thereby mapped to a simple dynamical system on the set of these 
dressed vacua.  A higher dimensional analogue of this hierarchy 
(an elliptic analogue of the Bogomolny hierarchy) is also presented.  
\end{abstract}
\bigskip

\begin{flushleft}
Mathematics Subject Classification:  35Q58, 37K10, 58F07\\
Key words: soliton equation, elliptic curve, holomorphic bundle, 
Grassmann variety\\
Running head: Landau-Lifshitz hierarchy and Grassmann variety
\end{flushleft}
\bigskip

\begin{flushleft}
arXiv:nlin.SI/0312002
\end{flushleft}
\newpage

%%%%%%%%%%%%%%%
%% main text %% 
%%%%%%%%%%%%%%%

\section{Introduction}

The Landau-Lifshitz equation 
\beqnn
  \rd_t \mathbf{S} 
  = \mathbf{S} \times \rd_x^2\mathbf{S} 
  + \mathbf{S} \times J\mathbf{S} 
\eeqnn
is the equation of motion of classical spin fields 
$\mathbf{S} = \tp{(S_1,S_2,S_3)}$ with totally anisotropic 
coupling constants $J = \diag(J_1,J_2,J_3)$, 
$J_1 \not= J_2 \not= J_3 \not= J_1$. This is 
an example of soliton equations whose zero-curvature 
representation is related to an elliptic curve. 
The building blocks $A(z)$ and $B(z)$ of its 
zero-curvature equation \cite{bib:Sk79,bib:Ch81} 
\beqnn
  [\rd_x - A(z),\; \rd_t - B(z)] = 0
\eeqnn
are indeed $2 \times 2$ matrices of elliptic functions 
of the spectral parameter $z$.  

It is now widely known, after the seminal work of 
Sato \cite{bib:SS82} and Segal and Wilson \cite{bib:SW85}, 
that infinite dimensional Grassmann varieties provide 
a universal language for understanding soliton equations.  
According to their observation, many soliton equations 
can be translated to a simple dynamical system on 
a subset of an infinite dimensional ``universal'' 
Grassmann variety.  This fundamental observation 
has been confirmed for a varieties of cases, including 
higher dimensional generalizations as well \cite{bib:Ta84}.  
Almost all of the cases thus examined, however, are equations 
with a {\it rational} zero-curvature representation, namely, 
those with rational matrices $A(z)$ and $B(z)$ of the spectral 
parameter $z$.  The status of soliton equations with an elliptic 
(and higher genus) spectral parameter in the language of 
Grassmann varieties still remains obscure.  This is the problem 
that we address in this paper.  

There are a few notable studies closely related to this issue.  
One is the work of Date et al. \cite{bib:DJKM83} on a free fermion 
formalism of the Landau-Lifshitz equation.  Although one should, 
in principle, be able to translate such a free fermion formalism 
to a Grassmannian picture based on the vector space of 
creation/annihilation operators, this is actually not an easy task.  
The work of Carey et al. \cite{bib:CHMS93} is more close to 
our standpoint.  They use an infinite dimensional Grassmann 
variety as a tool to analyze a factorization problem in 
a loop group, thereby solving the Landau-Lifshitz equation 
in much the same way as other soliton equations with 
a rational zero-curvature representation.  From our point of 
view, however, their approach is yet unsatisfactory, because 
their usage of the infinite dimensional Grassmann variety 
fails to incorporate a fundamental geometric structure of 
the Landau-Lifshitz equation.  

The geometric structure, also discussed by Carey et al., 
is hidden in the twisted double periodicity 
\beqnn
  A(z + 2\omega_a) = \sigma_a A(z) \sigma_a, \quad 
  B(z + 2\omega_a) = \sigma_a B(z) \sigma_a, \quad 
  a = 1,2,3, 
\eeqnn
of the matrices $A(z)$ and $B(z)$.  Here $z$ is 
the complex coordinate of the torus 
$\Gamma = \CC/(2\omega_1\ZZ + 2\omega_3\ZZ)$ 
that represents a nonsingular complex elliptic curve. 
$\omega_a$'s denote the half periods chosen to satisfy 
the linear relation $\omega_1 + \omega_2 + \omega_3 = 0$, 
and $\sigma_a$'s the Pauli matrices 
\beqnn
    \sigma_1 = \left(\begin{array}{cc}
                     0 & 1 \\
                     1 & 0 
               \end{array}\right), \quad 
    \sigma_2 = \left(\begin{array}{cc}
                     0 & -i \\
                     i & 0 
               \end{array}\right), \quad 
    \sigma_3 = \left(\begin{array}{cc}
                     1 & 0 \\
                     0 & -1 
               \end{array}\right). 
\eeqnn
Geometrically, this twisted double periodicity is related 
to a nontrivial holomorphic $\mathrm{sl}(2,\CC)$ bundle 
over the torus; the factorization method works because 
this bundle is rigid  \cite{bib:HM01}.  The same bundle is 
known to play a fundamental role in the elliptic Gaudin model 
and an associated conformal field theory as well \cite{bib:KT97}.  

Our strategy is, firstly, to construct an infinite dimensional 
Grassmann variety from a vector space $V$ of $2 \times 2$ 
matrices --- rather than two dimensional vectors as used 
by Carey et al. --- of Laurent series of $z$.  Secondly, 
we choose a special base point $W_0$ of the Grassmann variety. 
This base point has to be chosen so that the structure of 
the aforementioned $\mathrm{sl}(2,\CC)$ bundle is encoded 
therein.  More precisely, $W_0$ is to be identified with 
the space of its holomorphic sections over the punctured 
torus $\Gamma \setminus \{P_0\}$, where $P_0$ is the point 
at $z = 0$.  The Grassmann variety that fits this purpose 
turns out to be a slightly unusual one, which we call 
$\mathrm{Gr}_{-4}$.  The base point $W_0$, called ``vacuum'', 
is then ``dressed'' by a Lauernt series $\phi(z)$ to become 
a point of the Grassmann variety that corresponds to 
a general solution of the Landau-Lifshitz equation 
(or of an integrable hierarchy of evolution equations 
referred to as the Landau-Lifshitz hierarchy 
\cite{bib:GM91,bib:CHMS93}).  We thus construct a mapping 
to the set $\mathcal{M} \subset \mathrm{Gr}_{-4}$ of 
``dressed vacua.''  This mapping converts the Landau-Lifshitz 
equation (or hierarchy) to a simple dynamical system on 
$\mathcal{M}$.  

This paper is organized as follows.  Sections 2 and 3 
are mostly a review of basic notions.  Sections 2 is 
concerned with the factorization problem, and Section 3 
the construction of the Landau-Lifshitz hierarchy.  
New results are presented in Sections 4, 5 and 6.  
In Section 4, the infinite dimensional Grassmann variety, 
the set of dressed vacua and the mapping from 
the Landau-Lifshitz hierarchy are introduced.  
In Section 5, the dynamical system on the space of 
dressed vacua is specified.  Section 6 is a digression 
on a higher dimensional analogue of the Landau-Lifshitz 
hierarchy.  We conclude this paper with Section 7.

\section{Lie algebras and groups of Laurent series}

In the usual setting of a Riemann-Hilbert problem 
on the torus $\Gamma$ \cite{bib:Mi82,bib:Ro83}, 
one chooses a small circle $S^1 = \{z \mid |z| = a \}$ 
around the origin $z = 0$, and factorizes 
a (smooth or real analytic) map 
$g: S^1 \to \mathrm{SL}(2,\CC)$ to the product 
$g_{\mathrm{out}}(z)g_{\mathrm{in}}(z)$ of two holomorphic 
maps defined in the outside and inside of the circle.  
For an algebraic interpretation, however, it is simpler 
to avoid fixing the circle and to reorganize the factorization 
as follows \cite{bib:RS86}.  

Let $\fkg$ be the Lie algebra of Laurent series 
\beqnn
  X(z) = \sum_{n=-\infty}^\infty X_nz^n, \quad 
  X_n \in \mathrm{sl}(2,\CC), 
\eeqnn
that converges in a neighborhood of $z = 0$ except at $z = 0$. 
This Lie algebra has a direct sum decomposition of the form 
\beq
  \fkg = \fkg_{\mathrm{out}} \oplus \fkg_{\mathrm{in}}, 
\eeq
where $\fkg_{\mathrm{out}}$ and $\fkg_{\mathrm{in}}$ 
denote the following subalgebras: 
\begin{enumerate}
\item $\fkg_{\mathrm{in}}$ is the Lie subalgebra of 
all $X(z) \in \fkg$ that are also holomorphic at $z = 0$; 
in other words, 
\beq
  \fkg_{\mathrm{in}} = \{X(z) \in \fkg \mid 
     \mbox{$X_n = 0$ for $n < 0$}\}. 
\eeq
\item $\fkg_{\mathrm{out}}$ consists of all $X(z) \in \fkg$ 
that can be extended to a holomorphic mapping 
$X: \CC \setminus (2\omega_1\ZZ + 2\omega_3\ZZ) 
\to \mathrm{sl}(2,\CC)$ with singularity at each point of 
$2\omega_1\ZZ + 2\omega_3\ZZ$ and satisfy 
the twisted double periodicity condition 
\beq
  X(z + 2\omega_a) = \sigma_a X(z) \sigma_a 
  \quad a = 1,2,3. 
\eeq
\end{enumerate}
Note that constant matrices are excluded from 
$\fkg_{\mathrm{out}}$, so that 
$\fkg_{\mathrm{out}} \cap \fkg_{\mathrm{in}} = \{0\}$. 

One can use the well known weight functions 
\beq
  w_1(z) = \frac{\alpha\mathrm{cn}(\alpha z)}{\mathrm{sn}(\alpha z)}, 
  \quad 
  w_2(z) = \frac{\alpha\mathrm{dn}(\alpha z)}{\mathrm{sn}(\alpha z)}, 
  \quad 
  w_3(z) = \frac{\alpha}{\mathrm{sn}(\alpha z)}, 
\eeq
($\alpha = \sqrt{e_1 - e_3}$, $e_a = \wp(\omega_a)$, 
and $\mathrm{sn},\mathrm{cn},\mathrm{dn}$ are Jacobi's 
elliptic functions) in the zero-curvature representation 
\cite{bib:Sk79,bib:Ch81} to obtain a basis 
$\{\rd_z^nw_a(z)\sigma_a \mid n \ge 0, \; a = 1,2,3\}$ 
of $\fkg_{\mathrm{out}}$.   The projection 
$(\cdot)_{\mathrm{out}}: \fkg \to \fkg_{\mathrm{out}}$  
takes the simple form 
\beq
  \left(z^{-n-1}\sigma_a\right)_{\mathrm{out}} 
  = \frac{(-1)^n}{n!}\rd_z^nw_a(z) \sigma_a, \quad 
  \left(z^n\sigma_a\right)_{\mathrm{out}} = 0 \quad
  (n \ge 0) 
\eeq
in this basis.  

The direct sum decomposition of the Lie algebra $\fkg$ 
induces the factorization of the associated Lie group 
$G = \exp\fkg$ to the subgroups 
$G_{\mathrm{out}} = \exp\fkg_{\mathrm{out}}$ and 
$G_{\mathrm{in}} = \exp\fkg_{\mathrm{in}}$, 
namely, any element $g(z)$ of $G$ near the identity element $I$  
can be uniquely factorized as 
\beq
  g(z) = g_{\mathrm{out}}(z)g_{\mathrm{in}}(z), \quad 
  g_{\mathrm{out}}(z) \in G_{\mathrm{out}}, \quad 
  g_{\mathrm{in}}(z) \in G_{\mathrm{in}}. 
\eeq

\section{Construction of Landau-Lifshitz hierarchy}

The Landau-Lifshitz hierarchy is a multi-time nonlinear 
dynamical system on $G_{\mathrm{in}}$ derived from 
the linear dynamical system 
\beq
  g(z) \mapsto 
  g(z)\exp\Bigl(- \sum_{n=1}^\infty t_nz^{-n}\sigma_3\Bigr)
\eeq
on $G$ by the factorization \cite{bib:GM91}. 
Here $t = (t_1,t_2,\ldots)$ are the ``time'' variables 
of this system;  the first variable $t_1$ will be 
identified with the {\it spatial} variable of 
the Landau-Lifshitz equation in $1 + 1$ dimensions.  
The fundamental dynamical variable of this system 
is thus an $\mathrm{SL(2,\CC)}$-valued Laurent series 
of the form 
\beqnn
  \phi(z) = \sum_{n=0}^\infty \phi_nz^n, \quad 
  \det \phi(z) = 1, 
\eeqnn
that converges in a neighborhood of $z = 0$.  
The time evolution $\phi(0,z) \mapsto \phi(t,z)$ 
is achieved by the factorization 
\beq
  \phi(0,z)\exp\Bigl(- \sum_{n=1}^\infty t_nz^{-n}\sigma_3\Bigl) 
  = \chi(t,z)^{-1}\phi(t,z), 
  \label{factor-phi-exp}
\eeq
where $\chi(t,z)$ is an element of $G_{\mathrm{out}}$ 
that also depends on $t$.  

Equations of motion of $\phi(t,z)$ can be derived 
by the following standard procedure.  
Rewrite the factorization relation as 
\beqnn
  \chi(t,z)\phi(0,z)
  = \phi(t,z)\exp\Bigl(\sum_{n=1}^\infty t_nz^{-n}\sigma_3\Bigr), 
\eeqnn
differentiate both hand side by $t_n$, and 
eliminate $\phi(0,z)$ and the exponential matrix 
in the outcome by the factorization relation itself. 
This yields the relation 
\beq 
  \rd_{t_n}\chi(t,z)\cdot\chi(t,z)^{-1} 
  = \rd_{t_n}\phi(t,z)\cdot\phi(t,z)^{-1}
    + \phi(t,z)z^{-n}\sigma_3\phi(t,z)^{-1}. 
  \label{eq:log-der-chi-phi}
\eeq
Let $A_n(t,z)$ denote the $2 \times 2$ matrix 
defined by both hand sides of this relation.  
$A_n(t,z)$ thus has two expressions 
\beqnn
  A_n(t,z) = \rd_{t_n}\chi(t,z)\cdot\chi(t,z)^{-1} 
\eeqnn
and 
\beqnn
  A_n(t,z) 
  = \rd_{t_n}\phi(t,z)\cdot\phi(t,z)^{-1}
    + \phi(t,z)z^{-n}\sigma_3\phi(t,z)^{-1}. 
\eeqnn
The first expression shows that $A_n(t,z)$ takes 
values in $\fkg_{\mathrm{out}}$.  This implies 
that $A_n(t,z)$ should be equal to the projection, 
down to $\fkg_{\mathrm{out}}$, of the right hand side 
of the second expression.  Since the first term 
$\rd_{t_n}\phi(t,z)\cdot\phi(t,z)^{-1}$ on 
the right hand side disappears by the projection, 
one obtains the formula 
\beq
  A_n(t,z) = \Bigl(\phi(t,z)z^{-n}\sigma_3\phi(t,z)^{-1}
             \Bigr)_{\mathrm{out}}. 
\eeq
One can insert this formula into the linear equations 
\beq
  \rd_{t_n}\phi(t,z) 
  = A_n(t,z)\phi(t,z) - \phi(t,z)z^{-n}\sigma_3 
\eeq
(which follow from one of the foregoing expressions 
of $A_n(t,z)$) to obtain the nonlinear equations 
\beq
  \rd_{t_n}\phi(t,z) 
  = - \Bigl(\phi(t,z)z^{-n}\sigma_3\phi(t,z)^{-1}
      \Bigr)_{\mathrm{in}}\phi(t,z), 
\eeq
where $(\cdot)_{\mathrm{in}}$ stands for 
the projection $\fkg \to \fkg_{\mathrm{in}}$.  
This is the final form of equations of motion 
of $\phi(t,z)$.  

The zero-curvature equations 
\beq
  [\rd_{t_m} - A_m(t,z),\; \rd_{t_n} - A_n(t,z)] = 0 
\eeq
follow from the auxiliary linear system 
\beq
  (\rd_{t_n} - A_n(t,z))\chi(t,z) = 0 
\eeq
as the Frobenius integrability condition.  
Also note that 
\beq
  \psi(t,z) 
  = \phi(t,z)\exp\Bigl(\sum_{n=1}^\infty t_nz^{-n}\sigma_3\Bigr) 
\eeq
satisfies a linear system of the same form. 
The zero-curvature equation for $m = 1$ and $n = 2$ 
amounts to the Landau-Lifshitz equation;   
$t_1$ and $t_2$ are identified with the spatial 
coordinate $x$ and the time variable $t$ therein.  
The spin variables can be read off, in the matrix 
form $S = \sum_{a=1}^3 S_a\sigma_a$, from 
the Laurent expansion 
\beq
  A_1(z) = Sz^{-1} + O(1) 
\eeq
at $z = 0$ or, equivalently, 
\beq
  S = \phi(t,0)\sigma_3\phi(t,0)^{-1}. 
\eeq

\section{Infinite dimensional Grassmann variety} 

The construction of an infinite dimensional Grassmann variety 
starts with the choice of an infinite dimensional 
vector space $V$.  Two options are available here, 
namely, Sato's algebraic or complex analytic construction 
based on a vector space of Laurent series \cite{bib:SS82} 
and Segal and Wilson's functional analytic construction 
based on the Hilbert space of square-integrable functions 
on a circle \cite{bib:SW85}.  Of course, the latter will 
be a natural choice for the present setting.  

Let $V$ be the vector space of Laurent series 
\beqnn
  X(z) = \sum_{n=-\infty}^\infty X_nz^n, \quad 
  X_n \in \mathrm{gl}(2,\CC), 
\eeqnn
that converges in a neighborhood of $z = 0$ except at $z = 0$.  
Note that the coefficients are now taken from $\mathrm{gl}(2,\CC)$ 
i.e., complex $2 \times 2$ matrices without any algebraic 
constraints.  This vector space is a matrix analogue of 
$V^{\mathrm{ana}(\infty)}$ in Sato's list of models 
\cite{bib:SS82};  as noted therein, one can introduce 
a natural linear topology in this vector space.   
The following infinite dimensional 
Grassmann variety $\mathrm{Gr}_{-4}$ of closed 
vector subspaces $W \subset V$ turns out to be 
relevant to the Landau-Lifshitz hierarchy: 
\beq
  \mathrm{Gr}_{-4} 
  = \{ W \subset V \mid 
       \dim\Ker(W \to V/V_{+}) 
       = \dim\Coker(W \to V/V_{+}) - 4 < \infty \}.
\eeq
Here $\mathrm{V}_{+}$ denotes the subspace of 
all elements $X(z)$ of $V$ that are holomorphic 
and vanish at $z = 0$, i.e., 
\beq
  V_{+} = \{X(z) \in V \mid  \mbox{$X_n = 0$ for $n \le 0$}\}. 
\eeq
The map $W \to V/V_{+}$  is the composition of 
the inclusion $W \hookrightarrow V$ and 
the canonical projection $V \to V/V_{+}$.  

The next task is to choose a suitable base point, 
to be called ``vacuum,'' in this Grassmann variety;  
this base point should correspond to 
the ``vacuum solution'' $\phi(t,z) = I$ of 
the Landau-Lifshitz hierarchy.   A correct choice 
is the following vector subspace $W_0 \subset V$: 
\beq
  W_0 
  &=& \{X(z) \in V \mid 
  \mbox{$X(z)$ can be extended to a holomorphic mapping}
  \nonumber\\
  && \mbox{$X:\CC \setminus (2\omega_1\ZZ + 2\omega_3\ZZ) 
\to \mathrm{gl}(2,\CC)$ with the twisted double periodicity}
  \nonumber\\ 
  && X(z + 2\omega_a) = \sigma_a X(z) \sigma_a \;(a = 1,2,3)\}  
\eeq
This resembles the definition of $\fkg_{\mathrm{out}}$; 
the difference is, firstly, that $X(z)$ now takes values 
in $\mathrm{gl}(2,\CC)$ rather than $\mathrm{sl}(2,\CC)$, 
and secondly, that $X(z)$ can be a constant matrix.  
The following lemma implies that $W_0$ is indeed 
an element of $\mathrm{Gr}_{-4}$.  

\begin{lemma}
The linear mapping $W_0 \to V/V_{+}$ has the following 
properties: 
\begin{enumerate}
\item $\Ker(W_0 \to V/V_{+}) = \{0\}$. 
\item $\Im(W_0 \to V/V_{+}) \oplus \mathrm{sl}(2,\CC) 
\oplus \CC z^{-1}I = V/V_{+}$. 
\end{enumerate}
Here $\mathrm{sl}(2,\CC)$ and $\CC z^{-1}I$ are both 
understood to be embedded in $V/V_{+}$.  
\end{lemma}

\proof 
Notice that $\Ker(W_0 \to V/V_{+}) = W_0 \cap V_{+}$.  
If an element $X(z)$ of $W_0$ belongs to $V_{+}$, 
it is a matrix-valued holomorphic function defined 
everywhere on $\CC$ and has a zero at $z =0$.  
The twisted double periodicity implies that 
$X(z)$ is bounded.  By Liouville's theorem, 
such a function should be identically zero.  
To confirm the statement on $\Im(W_0 \to V/V_{+})$, 
notice that $W_0$ contains $I$, $\rd_z^nw_a(z)\sigma_a$ 
and $\rd_z^n\wp(z)I$ for $n \ge 0$ and $a = 1,2,3$.  
The latter two have the Laurent expansion 
\beq
  \rd_z^nw_a(z)\sigma_a = (-1)^n n!z^{-n-1}\sigma_a + O(z) 
\eeq
and 
\beq
  \rd_z^n\wp(z)I = (-1)^n (n+1)!z^{-n-2}I + O(z) 
\eeq
at $z = 0$.  One can thus see that the image of $W_0$ 
in $V/V_{+}$ contains $I$, $z^{-n-1}\sigma_a$ and 
$z^{-n-2}I$ for $n \ge 0$ among the standard basis 
$\{z^{-n}\sigma_a,\, z^{-n}I \mid n \ge 0, \; a = 1,2,3\}$ 
of $V/V_{+}$.  What is missing are 
$\sigma_1,\sigma_2,\sigma_3$ and $z^{-1}I$ 
that span the four dimensional subspace 
$\mathrm{sl}(2,\CC) \oplus \CC z^{-1}I$ of $V/V_{+}$.  
On the other hand, it is an easy exercise of 
complex analysis to shows that there is no element of 
$W_0$ that behaves as $\sigma_a + O(z)$, $a = 1,2,3$, 
or $z^{-1}I + O(z)$ at $z = 0$.  One can thus confirm 
that the image of $W_0$ in $V/V_{+}$ is complementary 
to $\mathrm{sl}(2,\CC) \oplus \CC z^{-1}I$.  
\qed

Stated differently, the lemma says that the composition 
of the inclusion $W_0 \hookrightarrow V$ and the canonical 
projection $V \to V/(V_{+}\oplus\mathrm{sl}(2,\CC)\oplus
\CC z^{-1}I)$ is an isomorphism 
\beq
  W_0 \simeq V/(V_{+}\oplus\mathrm{sl}(2,\CC)\oplus\CC z^{-1}I), 
\eeq
where $\mathrm{sl}(2,\CC)$ and $\CC z^{-1}I$ 
are now understood to be a subspace of $V$.  
This condition satisfied by $W_0$ is 
an {\it open condition}, namely, 
\beq
  \mathrm{Gr}_{-4}^\circ 
  = \{ W \in \mathrm{Gr}_{-4} \mid 
  W \simeq V/(V_{+}\oplus\mathrm{sl}(2,\CC)\oplus\CC z^{-1}I) \} 
\eeq
is an open subset of $\mathrm{Gr}_{-4}$, 
in fact, the open cell (or ``big cell'') 
in a cell decomposition of this Grassmann variety.  
Therefore, if a general solution $\phi(t,z)$ 
of the Landau-Lifshitz hierarchy is a small 
deformation of the vacuum solution $\phi(t,z) = I$, 
the ``dressed vacuum''  
\beq
  W(t) = W_0 \phi(t,z) 
\eeq
remains in this open subset.  The dynamics of 
the Landau-Lifshitz hierarchy is thus encoded 
to the motion of this dressed vacuum.

\section{Dynamical system in Grassmann variety} 

The consideration in the following is limited 
to a small deformation of the vacuum solution 
and small values of $t$.  The subspace $W(t) 
= W_0\phi(t,z)$ of $V$ thereby remains 
in the open cell $\mathrm{Gr}_{-4}^\circ$ of 
$\mathrm{Gr}_{-4}$.   

The dynamics of $W(t)$  turns out to take a simple form.  
This can be deduced from the factorization relation 
(\ref{factor-phi-exp}).  A clue is the following.  

\begin{lemma}
$W_0\chi(t,z) = W_0$.  
\end{lemma}

\proof 
By construction, $\chi(t,z)$ itself is an element 
of $W_0$.  $W_0$ is closed under multiplication 
of two elements, because the analytical properties 
in the definition of $W_0$ are preserved under multiplication.  
Consequently, $W_0\chi(t,z) \subseteq W_0$.  On the other hand, 
$\chi(t,z)$ is invertible, and the inverse matrix 
$\chi(t,z)^{-1}$ has the same analytical properties 
as $\chi(t,z)$.  Consequently, $\chi(t,z)^{-1}$ 
belongs to $W_0$, so that $W_0\chi(t,z)^{-1} \subseteq W_0$.  
This implies that the equality holds.  \qed

If one rewrites (\ref{factor-phi-exp}) as 
\beqnn
  \phi(t,z) 
  = \chi(t,z)\phi(0,z)
    \exp\Bigl(- \sum_{n=1}^\infty t_nz^{-n}\sigma_3\Bigr) 
\eeqnn
and apply this expression of $\phi(t,z)$ to 
the definition of $W(t)$, the outcome is that 
\beqnn
  W(t) 
  &=& W_0\chi(t,z)\phi(0,z)
      \exp\Bigl(- \sum_{n=1}^\infty t_nz^{-n}\sigma_3\Bigr) \\ 
  &=& W_0\phi(0,z)
      \exp\Bigl(- \sum_{n=1}^\infty t_nz^{-n}\sigma_3\Bigr) \\ 
  &=& W(0)\exp\Bigl(- \sum_{n=1}^\infty t_nz^{-n}\sigma_3\Bigr). 
\eeqnn
Note that the foregoing lemma has been used between 
the first and second lines of this calculation.  
Thus the following result has been deduced.  

\begin{theorem}
The Landau-Lifshitz hierarchy can be mapped, 
by the correspondence $\phi(t,z) \mapsto 
W(t) = W_0\phi(t,z)$, to a dynamical system on 
the subset 
\beq
   \mathcal{M} = \{W \in \mathrm{Gr}_{-4}^0 \mid 
      W = W_0\phi(z), \; \phi(z) \in G_{\mathrm{in}}\} 
\eeq
of $\mathrm{Gr}_{-4}$ with the exponential flows 
\beq
  W(t) 
= W(0)\exp\Bigl(- \sum_{n=1}^\infty t_nz^{-n}\sigma_3\Bigr). 
  \label{eq:exp-flow} 
\eeq
\end{theorem}

It will be instructive to derive, conversely, 
the factorization relation (\ref{factor-phi-exp}) 
from the dynamical system in Grassmann variety.  
Suppose that one is given an element $\phi(z)$ 
of $G_{\mathrm{in}}$ such that $W(0) = W_0\phi(z)$ 
belongs to the open cell $\mathrm{Gr}_{-4}^\circ$.  
The point $W(t)$ of the trajectory of 
the exponential flows (\ref{eq:exp-flow}) remains 
in the same open cell as far as $t$ is small.  
This implies that 
\beq
  \dim\Im(W(t) \to V/V_{+}) \cap \mathrm{gl}(2,\CC) = 1, 
\eeq
so that $\Im(W(t) \to V/V_{+}) \cap \mathrm{gl}(2,\CC)$ 
contains a nonzero matrix $\phi_0(t)$, which is close to 
the leading term of $\phi(z)$ if $t$ is sufficiently small.  
One can choose $\phi_0(t)$ to be unimodular, i.e., 
$\det\phi_0(t) = 1$.  Thus $W(t)$ turns out to contain 
an element of the form 
\beqnn
  \phi(t,z) = \sum_{n=0}^\infty \phi_n(t)z^n, \quad 
  \det\phi_0(t) = 1. 
\eeqnn
On the other hand, as an element of 
\beqnn
  W(t) 
  = W(0)\exp\Bigl(- \sum_{n=1}^\infty t_nz^{-n}\sigma_3\Bigr)
  = W_0\phi(z)\exp\Bigl(- \sum_{n=1}^\infty t_nz^{-n}\sigma_3\Bigr), 
\eeqnn
$\phi(t,z)$ can be written as 
\beqnn
  \phi(t,z) 
  = \chi(t,z)\phi(z)
    \exp\Bigl(- \sum_{n=1}^\infty t_nz^{-n}\sigma_3\Bigr) 
\eeqnn
with a $t$-dependent element $\chi(t,z)$ of $W_0$.  
Now examine the relation 
\beqnn
  \det\phi(t,z) = \det\chi(t,z) 
\eeqnn
that follows from the last equality.  Since $\chi(t,z)$ 
is an element of $W_0$, $\det\chi(t,z)$ becomes 
a doubly periodic holomorphic function of $z$ on 
$\CC \setminus (2\omega_1\ZZ + 2\omega_3\ZZ)$.  
On the other hand, one knows that $\det\phi(t,z) 
= 1 + O(z)$ as $z \to 0$, which implies that 
$\det\chi(t,z)$ is also holomorphic at each point 
of $2\omega_1\ZZ + 2\omega_3\ZZ$.  Hence, 
by Liouville's theorem, $\det\chi(t,z)$ turns out 
to be a constant.  Letting $z \to 0$, one finds that 
$\det\phi(t,z) = \det\chi(t,z) = 1$, so that 
\beqnn
  \chi(t,z) \in G_{\mathrm{out}}, \quad 
  \phi(t,z) \in G_{\mathrm{in}}. 
\eeqnn
One can thus see that the exponential flows 
(\ref{eq:exp-flow}) on the Grassmann variety 
solves the factorization problem 
(\ref{factor-phi-exp}).

\section{Elliptic analogue of Bogomolny hierarchy}

It is nowadays well known \cite{bib:MW96} that 
many $1 + 1$ dimensional soliton equations can be 
derived, via the Bogomolny equation in three dimensions, 
from the (anti)self-dual Yang-Mills equation in 
four dimensions.  Speaking differently, the latter 
may be thought of as a higher dimensional analogue 
of soliton equations.   

The problem addressed here is to construct a similar 
higher dimensional analogue of the Landau-Lifshitz hierarchy. 
Although the work of Carey et al. \cite{bib:CHMS93} 
briefly mentions a formulation of on such a higher dimensional 
analogue, this issue deserves to be investigated in more detail.  
For simplicity, the following consideration is limited to 
a higher dimensional analogue of the Bogomolny type;  
it is straightforward to generalize these results to 
equations of the self-duality type.  

A clue for constructing such a higher dimensional 
analogue is the fact that the subspace $W_0 \subset V$ 
is invariant, $W_0 f(z) \subseteq W_0$, under 
the multiplication by any elliptic function $f(z)$ 
with a pole at $z = 0$ and holomorphic elsewhere.   
In the context of the Landau-Lifshitz hierarchy, 
this implies that the one-parameter flow generated 
by $f(z)I$ on $\mathcal{M}$ is trivial: 
\beq
  W\exp(tf(z)I) = W 
  \quad (W \in \mathcal{M}). 
\eeq
Actually, as demonstrated in the usual cases 
(KdV, nonlinear Schr\"odinger, etc.) \cite{bib:IT01,bib:KIT02}, 
nontrivial higher dimensional flows stem from 
those trivial flows (or symmetries) of 
the soliton equations.  

Let $f_n(z)$, $n = 1,2,\ldots$, be a set of elliptic 
functions with a pole at $z = 0$ and holomorphic 
elsewhere, e.g., $f_n(z) = \rd_z^{n-1}\wp(z)$, 
and $s = (s_1,s_2,\ldots)$ a set of corresponding 
``time'' variables (though some of those may rather 
be understood as spatial variables in the context of 
the Bogomolny equation).  One needs yet another 
spatial variable $y$; a fundamental dynamical variable 
of the higher dimensional hierarchy is the Laurent series 
\beqnn
  \phi(y,z) = \sum_{n=0}^\infty \phi_n(y)z^n 
  \in G_{\mathrm{in}} 
\eeqnn
that depends on $y$.  In other words, the dynamical 
variable of this system is a $G_{\mathrm{in}}$-valued 
field on the $y$ space.  

The time evolution $\phi(0,y,z) \mapsto \phi(s,y,z)$ 
of this system is defined by the factorization relation 
\beq
  \phi\Bigl(0,\; y + \sum_{n=1}^\infty s_nf_n(z),\; z \Bigr)
  = \chi(s,y,z)^{-1}\phi(s,y,z), 
  \nonumber \\
  \chi(s,y,z) \in G_{\mathrm{out}}, \quad 
  \phi(s,y,z) \in G_{\mathrm{in}}. 
\eeq
In other words, this is the projection, down to 
$G_{\mathrm{in}}$, of the flows 
\beq
   g(y,z) \;\mapsto\; 
   g\Bigl(y + \sum_{n=1}^\infty s_nf_n(z),\; z \Bigr) 
   = \exp\Bigl(\sum_{n=1}^\infty s_nf_n(z)\rd_y\Bigr)g(y,z) 
\eeq
generated by an exponential operator on the space of 
$G$-valued fields with one spatial variable $y$.  

One can derive equations of motion of $\phi(s,y,z)$ 
in much the same way as the previous case.   
Firstly,  note that the differential operator 
$\rd_{s_n} - f_n(z)\rd_y$ annihilates the left hand side 
of the foregoing factorization relation.  
Applying this operator to both hand sides of 
the factorization relation and doing some algebra, 
one obtains the relation 
\beq
  (- \rd_{s_n} + f_n(z)\rd_y)\chi(s,y,z)\cdot\chi(s,y,z)^{-1} 
  = (- \rd_{s_n} + f_n(z)\rd_y)\phi(s,y,z)\cdot\phi(s,y,z)^{-1}, 
\eeq
where both hand sides are multiplied by $-1$ 
for convenience.  Let $B_n(s,y,z)$ denote the matrix 
defined by both hand sides of this relation.  
$B_n(s,y,z)$ thus has two expressions 
\beqnn
  B_n(s,y,z) 
  = - \rd_{s_n}\chi(s,y,z)\cdot\chi(s,y,z)^{-1} 
    + f_n(z)\rd_y\chi(s,y,z)\cdot\chi(s,y,z)^{-1}. 
\eeqnn
and 
\beqnn
  B_n(s,y,z) 
  = - \rd_{s_n}\phi(s,y,z)\cdot\phi(s,y,z)^{-1} 
    + f_n(z)\rd_y\phi(s,y,z)\cdot\phi(s,y,z)^{-1} 
\eeqnn
Notice here that $\rd_{s_n}\chi(s,y,z)\cdot\chi(s,y,z)^{-1}$ 
and $\rd_y\chi(s,y,z)\cdot\chi(s,y,z)^{-1}$, 
which appear in the first expression, take values 
in $\fkg_{\mathrm{out}}$.  Moreover, thanks to 
the analytic property of $f_n(z)$, the product 
of $f_n(z)$ and the second term remains in 
$\fkg_{\mathrm{out}}$.  Consequently, 
$B_n(s,y,z) \in \fkg_{\mathrm{out}}$.  
This implies that $B_n(s,y,z)$ should be equal 
to the projection, down to $\fkg_{\mathrm{out}}$, 
of the right hand side of the second expression 
of $B_n(s,y,z)$.  Since the first term 
$\rd_{s_n}\phi(s,y,z)\cdot\phi(s,y,z)^{-1}$ 
on the right hand side disappears by the projection, 
one obtains the formula 
\beq
  B_n(s,y,z) 
  = \Bigl(f_n(z)\rd_y\phi(s,y,z)\cdot\phi(s,y,z)^{-1} 
    \Bigr)_{\mathrm{out}}. 
\eeq
One can insert this formula into the linear equations 
\beq
  (\rd_{s_n} - f_n(z)\rd_y + B_n(s,y,z))\phi(s,y,z) = 0. 
\eeq
(which follow from one of the foregoing expressions 
of $B_n(s,y,z)$) to obtain a system of nonlinear 
equations for $\phi(s,y,z)$.  They give equations 
of motion of $\phi(s,y,z)$.  Moreover, 
the zero-curvature equations 
\beq
  [\rd_{s_m} - f_m(z)\rd_y + B_m(s,y,z),\; 
   \rd_{s_n} - f_n(z)\rd_y + B_n(s,y,z) ] = 0, 
\eeq
can be derived from the same linear equations or, 
equivalently, from 
\beq
  (\rd_{s_n} - f_n(z)\rd_y + B_n(s,y,z))\chi(s,y,z) = 0. 
\eeq
These equations comprise an elliptic analogue 
of the Bogomolny hierarchy.

\section{Conclusion}

We have elucidated the status of the Landau-Lifshitz 
hierarchy in the language of an infinite dimensional 
Grassmann variety.  The main result (Section 5) 
shows that the hierarchy can be mapped to a simple 
dynamical system on the set $\mathcal{M}$ of 
dressed vacua in the Grassmann variety $\mathrm{Gr}_{-4}$.  
The construction of the set of dressed vacua (in particular, 
the choice of the base point $W_0$) is closely related to 
the holomorphic $\mathrm{sl}(2,\CC)$ bundle hidden in 
the structure of the zero-curvature equations.  Thus 
the Landau-Lifshitz hierarchy turns out to fall into 
the universal framework of Sato \cite{bib:SS82} and 
Segal and Wilson \cite{bib:SW85}.  It is straightforward 
to generalize these results to the case of $\mathrm{sl}(N,\CC)$. 

Let us lastly stress an unusual nature of this case. 
Namely, we had to define the Grassmann variety itself 
in a slightly unusual way.  For most soliton equations, 
the relevant Grassmann variety consists of vector 
subspaces $W \subset V$ such that the linear map 
$W \to V/V_{+}$ is a Fredholm map of index zero; 
in our case, the index is equal to $-4$.  This is 
of course due to the special structure of $W_0$.

\subsection*{Acknowledgements}

I would like to thank Takeshi Ikeda and Takashi Takebe 
for comments and discussion on related issues.  
This work was partly supported by 
the Grant-in-Aid for Scientific Research (No. 14540172) 
from the Ministry of Education, Culture, 
Sports and Technology.


\begin{thebibliography}{99}


\bibitem{bib:CHMS93}
A.L. Carey, K.C. Hannabuss, L.J. Mason and M.A. Singer, 
The Landau-Lifshitz equation, elliptic curve, and 
the Ward transform, 
Commun. Math. Phys. {\bf 154} (1993), 25--47.  

\bibitem{bib:Ch81}
I.V. Cherednik, 
On integrability of the equation of a two-dimensional 
asymmetric $O(3)$-field and its quantum analogue, 
Yad. Fiz. {\bf 33} (1) (1981), 278--282.  

\bibitem{bib:DJKM83}
E. Date, M. Jimbo, M. Kashiwara and T. Miwa, 
Landau-Lifshitz equation: solitons, quasi-periodic 
solutions and infinite dimensional Lie algebras, 
J. Phys. A: Math. Gen. {\bf 16} (1983), 221--236.  

\bibitem{bib:GM91}
F. Guil and M. Ma\~{n}as, 
Loop algebras and the Krichever-Novikov equation, 
Phys. Lett. {153A} (1991), 90--94.  

\bibitem{bib:HM01}
J. C. Hurtubise and E. Markman, 
Surfaces and the Sklyanin bracket, 
Commun. Math. Phys. {\bf 230} (2002), 485--502.

\bibitem{bib:IT01}
T. Ikeda and K. Takasaki, 
Toroidal Lie algebras and Bogoyavlensky's 
2+1-dimensional equation, 
International Mathematics Research Notices {\bf 7} (2001), 
329--369. 

\bibitem{bib:KIT02} 
S. Kakei, T. Ikeda and K. Takasaki, 
Hierarchy of (2+1)-dimensional nonlinear Schr\"odinger equation,
self-dual Yang-Mills equation, and toroidal Lie algebras, 
Ann. Henri Poincare {\bf 3} (2002), 817--845.  

\bibitem{bib:KT97}
G. Kuroki and T. Takebe, 
Twisted Wess-Zumino-Witten models on elliptic curves, 
Commun. Math. Phys. {\bf 190} (1997), 1--56.

\bibitem{bib:MW96} 
L.J. Mason and N.M.J. Woodhouse, 
Integrability, self-duality and twistor theory 
(Clarendon Press, Oxford, 1996).  

\bibitem{bib:Mi82}
A.V. Mikhailov, 
The Landau-Lifshitz equation and the Riemann boundary problem 
on a torus, 
Phys. Lett. {\bf 92A} (1982), 51--55. 

\bibitem{bib:RS86}
A.G. Reyman and M.A. Semenov-Tian-Shansky, 
Zap. Nauchn. Sem. LOMI {\bf 150} (1986), 104--118; 
J. Soviet Math. {\bf 46} (1989), 1631--1640. 

\bibitem{bib:Ro83}
Yu.L. Rodin, 
The Riemann boundary problem on Riemann surfaces and 
the inverse scattering problem for the Landau-Lifshitz equation, 
Lett. Math. Phys. {\bf 7} (1983), 3--8. 
%% {\it ditto\/}, Physica {\bf 11D} (1984), 90--108. 

\bibitem{bib:SS82}
M. Sato and Y. Sato, 
{\it Soliton equations as dynamical systems on 
an infinite dimensional Grassmannian manifold\/}, 
Lecture Notes in Num. Appl. Anal., vol. 5, pp. 259--271 
(Kinokuniya, Tokyo, 1982). 

\bibitem{bib:SW85}
G.B. Segal and G. Wilson, 
{\it Loop groups and equations of KdV type\/}, 
Publ. Math. IHES {\bf 61} (1985), 5--65. 

\bibitem{bib:Sk79}
E.K. Sklyanin, 
On complete integrability of the Landau-Lifshitz equation, 
Steklov Mathematical Institute Leningrad Branch preprint 
LOMI, E-3-79, 1979. 

\bibitem{bib:Ta84}
K. Takasaki, 
A new approach to the self-dual Yang-Mills equations, 
Commun. Math. Phys. {\bf 94} (1984), 35--59. 

\end{thebibliography}
\end{document}